\documentclass[%
twocolumn,
groupedaddress,
superscriptaddress,
amsmath,amssymb,
ajp,
]{revtex4-2}

\usepackage{graphicx}
\usepackage{amsfonts}
\usepackage{bm}
\usepackage{hyperref,url}
\usepackage{changes}
\usepackage{xcolor}
\usepackage{soul}

\usepackage{ragged2e}

\begin{document}

\title{A geometrical-optics analogy for gravitational lensing using axicon-type lenses}

\author{Santiago Hern\'andez-D{\'\i}az}
\address{Institut f\"ur Astronomie und Astrophysik, Eberhard Karls Universit\"at\\ 
T\"ubingen, Sand 1, 72076 T\"ubingen, Germany}

\author{\'Angel S. Sanz}
\affiliation{Department of Optics, Faculty of Physical Sciences, Universidad Complutense de Madrid\\
Pza.\ Ciencias 1, Ciudad Universitaria E-28040 Madrid, Spain}


\begin{abstract}
Gravitational lensing is often introduced through the bending of light by mass, but its geometrical nature can be difficult to visualize without invoking the full machinery of general relativity. We present a geometrical-optics analogy in which selected lensing-like image morphologies are generated by ray tracing through axicon-type lenses. In contrast with analogies based on a spatially varying refractive index, the present model uses optical elements with a constant refractive index; the redistribution of rays is produced by the geometry of the refracting surfaces. After deriving the ray-tracing equations for a general two-surface axicon, we apply the model to conical and exponential profiles. Axially symmetric configurations generate ring-like images, misaligned incidence produces partial arcs, and broken axial symmetry leads to four-image patterns reminiscent of an Einstein cross.
The model is not intended as a physical substitute for a relativistic lens equation, but as a computational and pedagogical tool for exploring how surface geometry and symmetry control ray deflection, redistribution, and image multiplicity.
\end{abstract}

\maketitle


\section{Introduction}
\label{sec:introduction}

Gravitational lensing provides one of the most striking examples of the geometrical character of light propagation. In general relativity, light does not bend because it is refracted by a material medium, but because its trajectory follows null geodesics in a curved space-time. This idea is conceptually powerful, but it is also difficult to visualize at an introductory or intermediate level without appealing to the full mathematical structure of general relativity. For this reason, gravitational lensing is a useful topic in physics education: it connects geometrical optics, symmetry, image formation, and the role of geometry in physical theories.

The historical development of the subject is well known. The possibility that gravity could deflect light was already considered by Soldner~\cite{sun}, and later became one of the classic observational tests of general relativity after Einstein's prediction of the relativistic deflection angle and the eclipse measurements reported by Dyson, Eddington, and Davidson~\cite{eclipse}. Since then, gravitational lensing has developed into a major tool in astronomy and cosmology. It can be used, for example, to infer cosmological distances and the Hubble constant~\cite{Hubble}, to estimate the mass of galaxy clusters~\cite{cluster}, and to constrain the distribution of dark matter~\cite{darkmatter}. Comprehensive introductions to the subject can be found in Refs.~\cite{astronomy,book1}. For a guide to the gravitational-lensing literature aimed at the AJP readership, see the Resource Letter by Treu, Marshall, and Clowe~\cite{Treu2012}.

From a teaching perspective, however, the richness of gravitational lensing also presents a difficulty. The phenomenon is often introduced either through a relativistic description, which may be too advanced for many students, or through simplified lens equations whose geometrical content is not always immediately visible. Several approaches have therefore been developed to make gravitational lensing more accessible. Some are based on numerical ray tracing and image simulation, as in astrophysical codes designed to reproduce realistic lensing observations~\cite{SKYLENS,mesh,Treecode,noiseless}.
Others exploit optical analogies. In particular, one may describe the gravitational field in terms of an effective refractive index of the vacuum, which varies spatially under the influence of the gravitational field~\cite{vacuum}. Such approaches are valuable because they translate part of the relativistic problem into the language of geometrical optics.

A related tradition, particularly represented in this journal, uses optical devices or simplified geometrical-optics models to make gravitational lensing tangible in teaching contexts. An early example is the gravitational-lens simulator proposed by Liebes~\cite{Liebes1969}. Subsequent works described plastic lenses designed to mimic the deflection produced by static gravitational fields or by selected mass distributions~\cite{Higbie1981,Adler1995}. At a more classroom-oriented level, Ros showed how gravitational-lensing-like effects can be demonstrated with simple transparent objects such as a drinking glass or its base~\cite{Ros2008}. More recently, Selmke introduced a controllable optical analogy based on liquid menisci for studying lensing by multi-component systems~\cite{Selmke2021}, while Szafraniec and Harford proposed a simple geometrical-optics model based on a spatially varying refractive index to illustrate Einstein rings and multiple images~\cite{Szafraniec2024}.

It is important to emphasize that the use of axicon-like optical elements in gravitational-lensing analogies is not new and has appeared in different pedagogical and experimental contexts~\cite{Liebes1969,Higbie1981,Adler1995,Selmke2021}. The present work does not aim to introduce axicons as gravitational-lens simulators. Instead, its purpose is to develop a compact and reproducible ray-tracing framework for two-surface axicon systems, with particular emphasis on how surface geometry and symmetry control the redistribution of rays and the resulting image morphologies. In contrast with approaches based on spatially varying refractive indices~\cite{vacuum,Szafraniec2024}, the refractive index here is kept constant within each optical region, so that the geometrical role of the refracting interfaces becomes fully explicit.

The model is therefore formulated as an explicit three-dimensional ray-tracing calculation suitable for advanced undergraduate study and computational projects. Its distinctive feature is that geometry, rather than a spatially varying optical medium, is the active ingredient: the redistribution of rays is controlled by the shape and symmetry of the refracting interfaces. This provides a transparent computational and pedagogical framework for investigating how surface geometry governs ray deflection and image multiplicity.

Axicons are especially suitable for this purpose. Since their introduction by McLeod~\cite{axicon}, axicons have been used as optical elements with unusual focusing properties. Unlike ordinary spherical lenses, an ideal axicon does not focus an on-axis point source into a single focal point, but into an extended region along the optical axis. This absence of a unique focal length makes axicons a natural candidate for constructing analogies with gravitational lenses, which also do not behave as ordinary thin lenses with a single focal point. Modern axicon designs and applications include a broad range of optical systems~\cite{applications}, and different refractive profiles, including conical and logarithmic forms, have been analyzed in the optics literature~\cite{three}. Axicons have also been considered in applied optical contexts such as free-space optical systems~\cite{FSO}.

The analogy developed below is based on elementary ingredients from geometrical optics. We model an axicon-type lens as an optical element of refractive index $n$ bounded by two surfaces, $z=f_1(x,y)$ and $z=f_2(x,y)$, embedded in an external medium of refractive index $n_0$.
Rays are propagated through the three regions separated by these surfaces by applying Snell's law at each interface.
The refracted directions are obtained in vector form from the local surface normals and the corresponding incidence conditions, leading to a compact three-dimensional ray-tracing algorithm.
The resulting ray distributions are then recorded on an observation plane and represented as smooth ray-density maps.

This work is organized as follows.
Within the above-mentioned framework, three familiar lensing-like morphologies are used as guiding examples: ring-like images from axial symmetry, partial arcs from tilted incidence, and four-image patterns from broken axial symmetry.
These examples set the stage for the more detailed discussion of the analogy in Sec.~\ref{sec:analogy}.
More specifically, this section discusses the scope of the analogy and clarifies the correspondence between gravitational-lensing features and the axicon-based optical model. Section~\ref{sec:raymodel} develops the ray-tracing model for a general two-surface axicon. Section~\ref{sec:profiles} introduces the conical and exponential profiles used in the simulations, including the symmetric and asymmetric configurations. Section~\ref{sec:results} presents the simulated ring-like, arc-like, and four-image patterns. Section~\ref{sec:pedagogy} discusses the pedagogical use of the analogy and its limitations. Finally, Sec.~\ref{sec:conclusions} summarizes the main conclusions.


\section{Gravitational lensing as a geometrical-optics analogy}
\label{sec:analogy}

The analogy proposed here is deliberately limited in scope. It does not attempt to reproduce the gravitational lens equation or to assign an optical refractive index to a gravitational field. Instead, it isolates one structural feature that is common to both systems: the redistribution of light rays is constrained by geometry. In gravitational lensing, the relevant geometry is that of space-time and light follows null geodesics. In the optical system considered here, the relevant geometry is that of the refracting surfaces and light is redirected according to Snell's law.

This distinction is important. A common way of connecting gravitational fields with ordinary optics is to introduce an effective refractive index for the vacuum, which varies with position and bends light rays in a way analogous to a graded-index medium~\cite{vacuum}. Such a description provides a useful bridge between general relativity and optical refraction. The analogy developed here is complementary: the refractive index is constant within each region, and the bending of rays arises exclusively from the shape of the interfaces. In this sense, the axicon surfaces do not represent the gravitational field itself. Rather, they provide a controllable geometrical device with which students can explore how symmetry and surface curvature affect ray deflection and image formation.

The analogy is guided by three qualitative features of strong gravitational lensing. The first is axial symmetry. When a source, a lens, and an observer are aligned, and the lensing mass distribution is sufficiently symmetric, the image of the source may form a ring. In the optical analogy, an axially symmetric axicon redistributes an initially circular bundle of rays into a ring-like pattern on an observation plane. The second feature is misalignment. If the source, lens, and observer are not perfectly aligned, the ring is broken into arcs. In the optical analogy, this situation is mimicked by tilting the incident ray bundle with respect to the symmetry axis of the axicon. The third feature is broken axial symmetry. In gravitational lensing, an elongated or otherwise non-axisymmetric mass distribution can produce several separated images of the same source. In the optical analogy, a similar qualitative effect is obtained by deforming the axicon profile so that the two transverse directions are no longer equivalent.

\begin{figure*}[t]
\begingroup 
\makeatletter
\long\def\@makecaption#1#2{%
 \par
 \vskip\abovecaptionskip
 \noindent
 \parbox{\linewidth}{%
 \small
 \setlength{\parindent}{0pt}%
 \setlength{\leftskip}{0pt}%
 \setlength{\rightskip}{0pt}%
 \setlength{\parfillskip}{0pt plus 1fil}%
 \setlength{\hangindent}{0pt}%
 \justifying
 \noindent#1.\enspace #2\par
 }%
 \vskip\belowcaptionskip
 }%
 \makeatother
 \centering
 \begin{minipage}[t]{0.43\textwidth}
 \centering
 \includegraphics[height=4.5cm, keepaspectratio ]{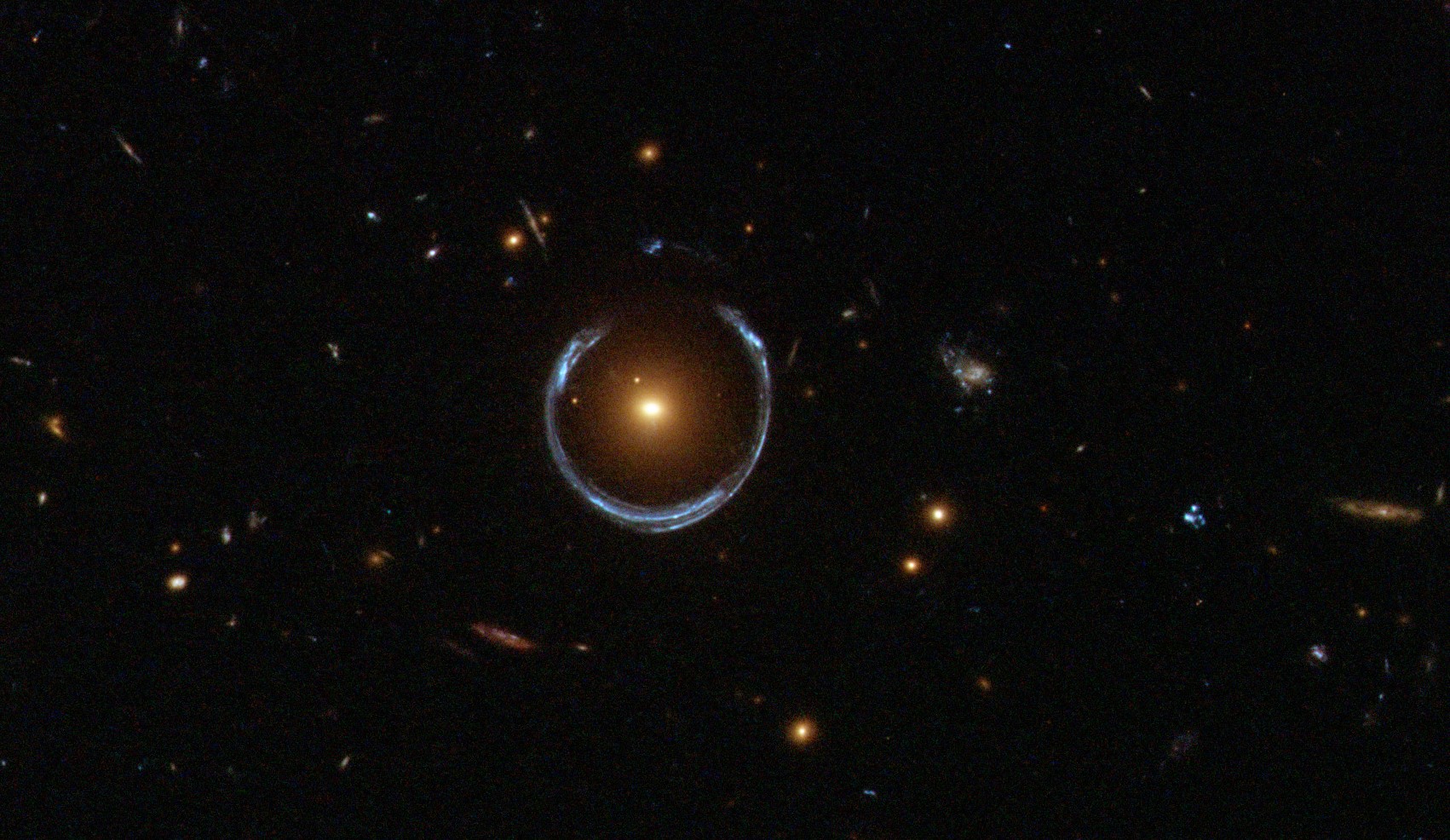} \vspace{2pt} \textbf{(a)}
 \end{minipage}
 \hfill
 \begin{minipage}[t]{0.25\textwidth}
 \centering
 \includegraphics[height=4.5cm, keepaspectratio ]{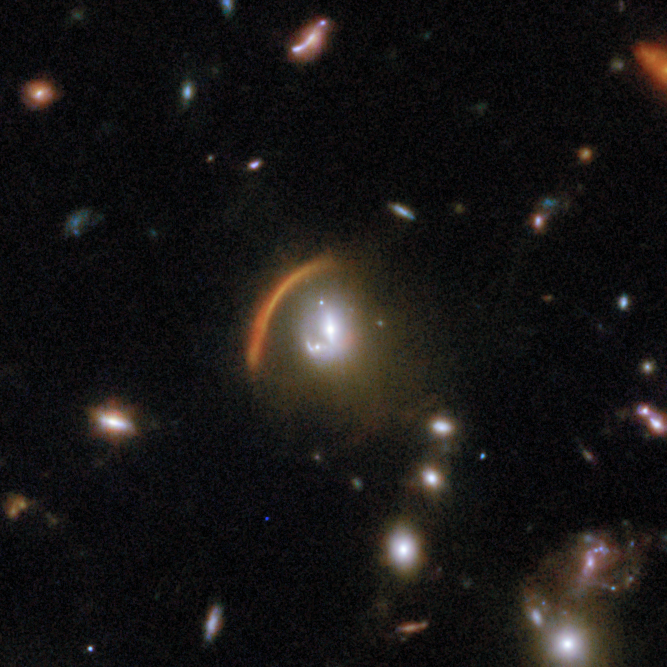} \vspace{2pt} \textbf{(b)}
 \end{minipage}
 \hfill
 \begin{minipage}[t]{0.27\textwidth}
 \centering
 \includegraphics[height=4.5cm, keepaspectratio ]{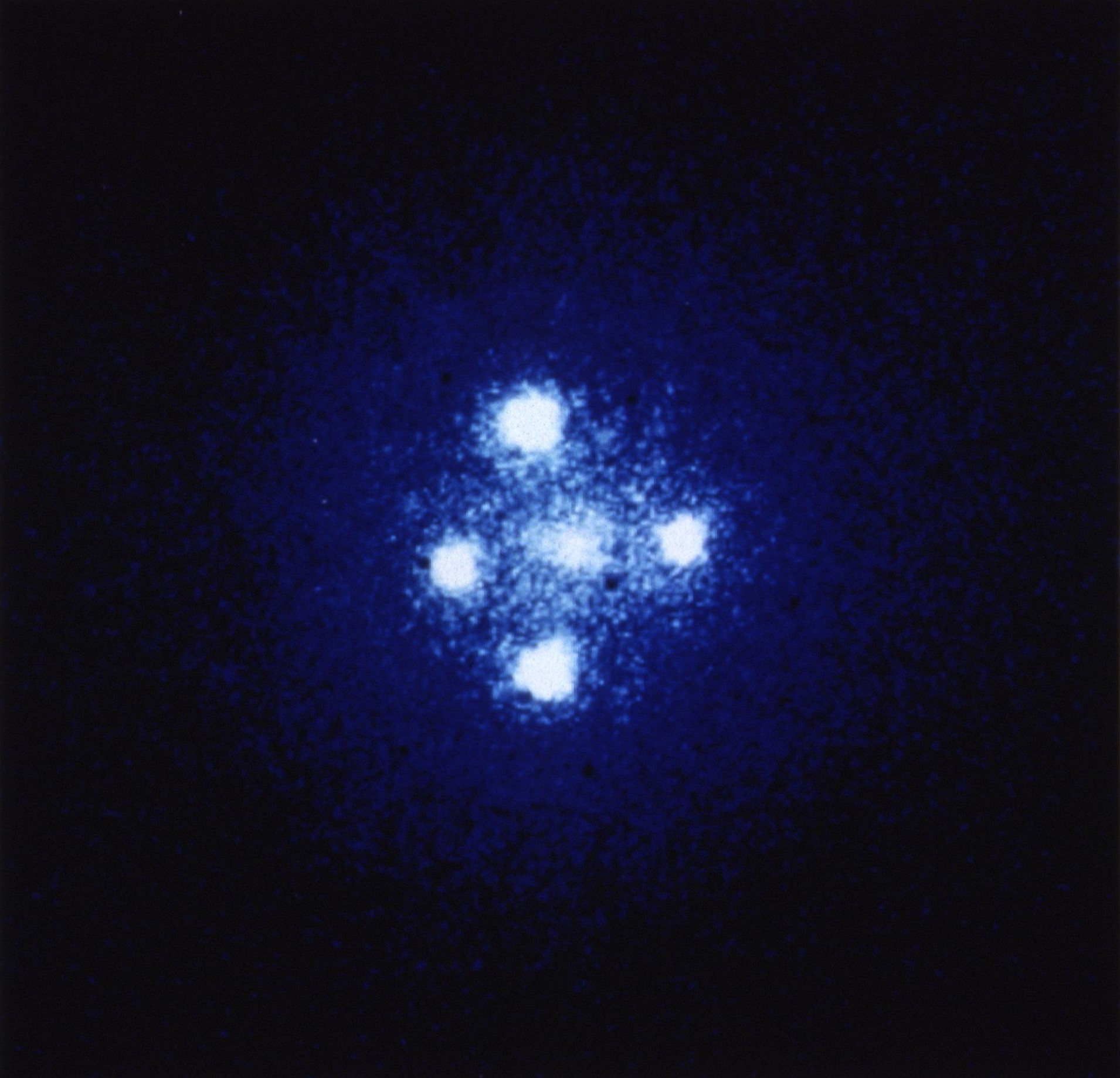} \vspace{2pt} \textbf{(c)}
 \end{minipage}
 \caption{Representative strong-gravitational-lensing morphologies. (a) The nearly ring-like image of the background galaxy in the system LRG~3-757~\cite{NASA_EinsteinRing}, also known as the Cosmic Horseshoe. (b) The gravitational lens COSJ100013+023424~\cite{ESA_GravitationalArc}, in which a distant galaxy appears as an extended arc beside the foreground lensing galaxy. (c) The Einstein Cross G2237+0305~\cite{NASA_EinsteinCross}, showing four images of a background quasar around the foreground lensing galaxy. The panels correspond qualitatively to the ring-like, arc-like, and four-image ray-density patterns considered in this work. They are not shown at a common angular scale, and no quantitative correspondence with the axicon parameters is implied. Image credits: (a) NASA and ESA; (b) ESA/Webb, NASA \& CSA, G.~Gozaliasl, A.~Koekemoer, and M.~Franco; (c) NASA, ESA, and STScI.}
 \label{fig:observational_examples}
 \endgroup
\end{figure*}

\begin{table*}[!t]
\caption{\label{tab:analogy}Qualitative correspondence between selected gravitational-lensing features and the axicon-based geometrical-optics analogy. The entries indicate the pedagogical role of the analogy and should not be interpreted as physical equivalences.}
\begin{ruledtabular}
\begin{tabular}{ll}
Gravitational-lensing feature & Axicon-based optical analogue \\
\hline
Curved space-time guiding null geodesics & Refracting surfaces redirecting rays by Snell's law \\
Axially symmetric lensing mass & Axially symmetric axicon profile \\
Einstein ring from alignment & Ring-like ray distribution on an observation plane \\
Source-lens-observer misalignment & Tilted incident ray bundle \\
Partial arcs & Broken ring-like ray-density pattern \\
Non-axisymmetric lensing mass & Asymmetric axicon profile \\
Multiple images / Einstein-cross morphology & Four separated ray-density maxima \\
\end{tabular}
\end{ruledtabular}
\end{table*}

Representative astronomical examples of these three morphologies are shown in Fig.~\ref{fig:observational_examples}. Panel (a) illustrates a nearly ring-like configuration produced by a high degree of alignment~\cite{NASA_EinsteinRing}, panel (b) shows an extended gravitational arc~\cite{ESA_GravitationalArc}, and panel (c) shows four images of a background quasar arranged around a foreground galaxy~\cite{NASA_EinsteinCross}.
Additional Einstein-ring observations obtained with the Hubble Space Telescope are collected in Ref.~\cite{NASA_EinsteinRingsGallery}.
These observations provide visual motivation for the optical patterns considered below. The comparison is strictly morphological: the simulations are not intended to reproduce the angular scales, brightness distributions, or lens parameters of these particular systems.

These correspondences are summarized in Table~\ref{tab:analogy}. The table should be read as a guide to the analogy, not as a dictionary between gravitational and optical quantities. In particular, the parameters of the axicon are not fitted to a mass distribution, and the observed patterns are not intended to reproduce the angular scales, magnifications, or time delays of a real lensing system. Their role is instead to make visible, within a familiar ray-optics framework, how geometrical constraints can transform a source distribution into rings, arcs, or multiple images.

The absence of a unique focal length is useful pedagogically because it prevents the analogy from being reduced to the familiar behavior of a thin lens. Instead, attention is directed to the redistribution of ray bundles. By changing the profile of the axicon, or by breaking its axial symmetry, one can generate different high-density ray patterns and discuss them in terms of elementary geometrical optics.

The ray-tracing model described in the next section provides the mathematical implementation of this analogy. It keeps the optical ingredients deliberately simple: piecewise constant refractive indices, two refracting surfaces, Snell's law, and propagation to a chosen observation plane. This simplicity is one of the advantages of the construction, since it allows the connection between surface geometry and image morphology to be followed step by step.


\section{Ray tracing through an axicon-type lens}
\label{sec:raymodel}

\begin{figure}[t]
 \centering
 \includegraphics[width=\columnwidth]{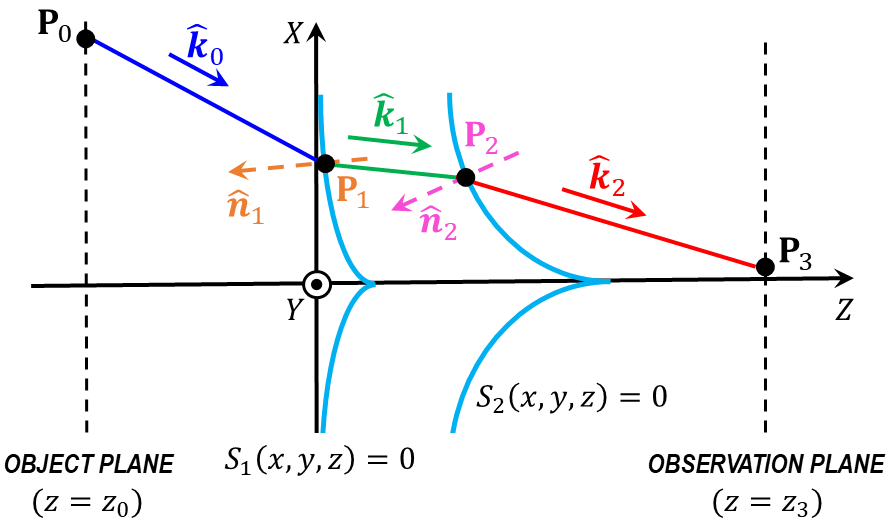}
 \caption{\label{fig1}
  Geometry of the ray-tracing model. 
  A ray launched from the point ${\bf P}_0$ at the object plane intersects the two refracting surfaces $S_1$ and $S_2$ at points ${\bf P}_1$ and ${\bf P}_2$, respectively, before reaching the observation plane at ${\bf P}_3$. 
  In the figure, $\hat{\bf k}_0$, $\hat{\bf k}_1$, and $\hat{\bf k}_2$ denote the unit direction vectors in the three propagation regions of interest, namely the incidence region ($0$), the interior of the axicon-type lens ($1$), and the transmission region ($2$), respectively.}
\end{figure}

We consider a generalized axicon-type optical element with homogeneous refractive index $n$, embedded in a homogeneous surrounding medium of refractive index $n_0$, as shown in Fig.~\ref{fig1}.
The first or front refracting surface $S_1$ is defined as the set of points ${\bf P}_1 = (x_1,y_1,z_1)$ satisfying
\begin{equation}
 S_1(x_1,y_1,z_1) = 0 .
\end{equation}
In the same manner, the second or back refracting surface $S_2$ is defined as
\begin{equation}
 S_2(x_2,y_2,z_2) = 0 ,
\end{equation}
for all points ${\bf P}_2 = (x_2,y_2,z_2)$.
These optical surfaces divide the ray path into three propagation regions: the incident region, the interior of the axicon, and the transmitted region.
The object plane is located at $z=z_0$, and the observation plane at $z=z_3$.

A ray is launched from an initial point
\begin{equation}
 {\bf P}_0=(x_0,y_0,z_0) ,
\end{equation}
at the object plane, with unit direction vector
\begin{equation}
 \hat{\bf k}_0 = (k_{0,x},k_{0,y},k_{0,z}) .
\end{equation}
Taking the optical axis ($z$ axis) as the polar axis of a spherical coordinate system, the propagation direction can be parameterized in terms of the polar and azimuthal angles $(\theta,\phi)$ as
\begin{equation}
  \hat{\bf k}_0 = (\sin\theta_0\cos\varphi_0,\sin\theta_0\sin\varphi_0,\cos\theta_0) ,
 \label{eq:k0}
\end{equation}
where $0\leq\theta_0<\pi/2$ and $0\leq\varphi_0<2\pi$.
Thus, in the incidence region, the path described by the ray coming from ${\bf P}_0$ can be parameterized in terms of a local path parameter $s$, measured from ${\bf P}_0$, as
\begin{equation}
 {\bf r}_0(s) = {\bf P}_0 + s \hat{\bf k}_0, \qquad 0\leq s\leq s_1 ,
 \label{eq:r0}
\end{equation}
where $s_1$ is defined through the point ${\bf P}_1$ where the ray intersects the first refracting surface,
\begin{equation}
 \begin{aligned}
 {\bf P}_1 & = (x_1,y_1,z_1) \\
 & \equiv {\bf r}_0(s_1) = (x_0(s_1),y_0(s_1),z_0(s_1)) \in S_1 ,
 \end{aligned}
\end{equation}
that is, any point for which the following condition is satisfied
\begin{equation}
 S_1(x,y,z) \arrowvert_{P_1} = S_1(x_1,y_1,z_1) = 0 .
 \label{eq:first_crossing}
\end{equation}

When the first surface is expressed explicitly as
\begin{equation}
 z=f_1(x,y) ,
 \label{eq:first_crossing3}
\end{equation}
the corresponding implicit surface function may be written as $S_1(x,y,z)=z-f_1(x,y)$, and the intersection condition in Eq.~\eqref{eq:first_crossing} becomes
\begin{equation}
 z_1=f_1(x_1,y_1).
 \label{eq:first_crossing2}
\end{equation}
Depending on the functional form of $f_1$, the corresponding value of $s_1$ may be obtained analytically or numerically. In the simulations below, the intersection equations are solved by bisection.
The functional form of $f_1(x,y)$ determines the local geometry of $S_1$ and, consequently, the surface normal at ${\bf P}_1$, given by the unit vector
\begin{equation}
 \hat{\bf m}_1 = \frac{\nabla S_1(x,y,z) \arrowvert_{P_1}}{\| \nabla S_1(x,y,z) \arrowvert_{P_1} \| }  .
 \label{eq:4}
\end{equation}
This construction applies at regular surface points for which $\nabla S_1\neq{\bf 0}$. The surface normal is not uniquely defined exactly at an idealized axicon apex; rays incident at that single point are therefore excluded from the calculation.
For refraction, it is common to orient the normal toward the incident medium.
We therefore define the normal unit vector as
\begin{equation}
    \hat{\bf n}_1=
    \begin{cases}
    \hat{\bf m}_1, & \hat{\bf k}_0\cdot\hat{\bf m}_1\leq 0,\\
    -\hat{\bf m}_1, & \hat{\bf k}_0\cdot\hat{\bf m}_1>0.
    \end{cases}
    \label{eq:n1}
\end{equation}

At the first interface, $\alpha_i$ and $\alpha_t$ denote the incidence and transmission angles, respectively; the corresponding angles at the second interface are denoted by $\beta_i$ and $\beta_t$.
Thus, with this convention, the physical incidence angle satisfies $0\leq\alpha_i\leq\pi/2$, which is defined by
\begin{equation}
 \cos\alpha_i=-\hat{\bf k}_0\cdot\hat{\bf n}_1 ,
 \label{eq:alphai}
\end{equation}
Snell's law at the first interface, where the ray passes from the external medium into the axicon, thus gives
\begin{equation}
 \sin\alpha_t = \frac{n_0}{n}\sin\alpha_i .
 \label{eq:alphat}
\end{equation}

If $\sin\alpha_i\neq 0$, the unit vector tangent to the ray in the plane of incidence is
\begin{equation}
 \hat{\boldsymbol{\tau}}_1 = \frac{\hat{\bf k}_0+\cos\alpha_i\,\hat{\bf n}_1}{\sin\alpha_i}.
    \label{eq:tau1}
\end{equation}
The refracted direction inside the axicon is then
\begin{equation}
 \hat{\bf k}_1 = -\cos\alpha_t\,\hat{\bf n}_1 + \sin\alpha_t\,\hat{\boldsymbol{\tau}}_1 ,
    \label{eq:k1}
\end{equation}
where
\begin{equation}
 \cos\alpha_t= \sqrt{1-\sin^2\alpha_t},
\end{equation}
where the positive root selects propagation into the second optical region.
Note that, for normal incidence on $S_1$, $\sin\alpha_i=0$, Eq.~\eqref{eq:k1} is replaced by the limiting result
\begin{equation}
  \hat{\bf k}_1=-\hat{\bf n}_1 ,
\end{equation}
in agreement with the sign convention \eqref{eq:n1}.

The ray trajectory inside the axicon is then described with a second local path parameter $u$, measured from ${\bf P}_1$:
\begin{equation}
    {\bf r}_1(u) = {\bf P}_1 + u \hat{\bf k}_1 , \qquad 0\leq u\leq u_2 ,
    \label{eq:r1}
\end{equation}
where the second intersection parameter $u_2$ is found from
\begin{equation}
 S_2(x_1(u_2),y_1(u_2),z_1(u_2)) = 0 ,
    \label{eq:u2}
\end{equation}
analogous to \eqref{eq:first_crossing}.
If the second surface is written explicitly as $z=f_2(x,y)$, then Eq.~\eqref{eq:u2} becomes
\begin{equation}
 z_2 = f_2(x_2,y_2) ,
 \label{eq:u22}
\end{equation}
in direct analogy with Eq.~\eqref{eq:first_crossing2}.
The solution provides us with the second intersection point
\begin{equation}
 {\bf P}_2 \equiv (x_2,y_2,z_2) = {\bf P}_1+u_2 \hat{\bf k}_1 .
    \label{eq:P2}
\end{equation}

The procedure follows as before, from Eq.~\eqref{eq:4} onward.
That is, we determine the normal direction to $S_2$ at ${\bf P}_2$ defining the auxiliary unit vector
\begin{equation}
 \hat{\bf m}_2 = \frac{\nabla S_2(x,y,z) \arrowvert_{P_2}}{\| \nabla S_2(x,y,z) \arrowvert_{P_2} \| }  .
 \label{eq:m2}
\end{equation}
Taking the same sign convention as before, we define the normal unit vector as
\begin{equation}
    \hat{\bf n}_2=
    \begin{cases}
    \hat{\bf m}_2, & \hat{\bf k}_1\cdot\hat{\bf m}_2\leq 0,\\
    -\hat{\bf m}_2, & \hat{\bf k}_1\cdot\hat{\bf m}_2>0.
    \end{cases}
    \label{eq:n2}
\end{equation}
The physical incidence angle at the second surface is therefore
\begin{equation}
 \cos\beta_i=-\hat{\bf k}_1\cdot\hat{\bf n}_2 ,
 \label{eq:betai}
\end{equation}
At this interface the ray passes from the axicon into the external medium, so Snell's law gives
\begin{equation}
 \sin\beta_t = \frac{n}{n_0}\sin\beta_i .
 \label{eq:betat}
\end{equation}
If $\sin\beta_t>1$, no real transmitted direction exists and total internal reflection occurs.

When no total internal reflection occurs and $\sin\beta_i\neq 0$, the tangent direction in the plane of incidence is
\begin{equation}
 \hat{\boldsymbol{\tau}}_2 = \frac{\hat{\bf k}_1 + \cos\beta_i\,\hat{\bf n}_2}
    {\sin\beta_i}.
    \label{eq:tau2}
\end{equation}
The transmitted direction in the external medium is then
\begin{equation}
 \hat{\bf k}_2 = -\cos\beta_t\,\hat{\bf n}_2 + \sin\beta_t\,\hat{\boldsymbol{\tau}}_2 .
 \label{eq:k2}
\end{equation}
For normal incidence, $\sin\beta_i=0$, Eq.~\eqref{eq:k2} is replaced by the limiting result
\begin{equation}
 \hat{\bf k}_2 = -\hat{\bf n}_2 .
\end{equation}
This vector form of Snell's law is equivalent to rotating the incident ray within the plane of incidence, but it avoids any ambiguity associated with the sign of the surface normal and remains well defined at normal incidence.

Finally, in the transmitted region we use a third local path parameter $v$, measured from ${\bf P}_2$:
\begin{equation}
 {\bf r}_2(v) = {\bf P}_2 + v\hat{\bf k}_2 , \qquad 0\leq v\leq v_3 .
 \label{eq:r2}
\end{equation}
The value $v_3$ at which the ray reaches the observation plane is determined by imposing
\begin{equation}
    z_2+k_{2,z}\,v_3=z_3,
\end{equation}
which gives
\begin{equation}
    v_3=\frac{z_3-z_2}{k_{2,z}},
    \label{eq:vo}
\end{equation}
provided that $k_{2,z}\neq 0$. The arrival point is therefore
\begin{equation}
    {\bf P}_3={\bf P}_2+v_3\hat{\bf k}_2,
    \label{eq:Po}
\end{equation}
with $z_3$ as its third coordinate. The coordinates $(x_3,y_3)$ are then used to build the ray-density map on the observation plane.

Repeating this procedure for a large number of initial points and incident directions produces a set of arrival points on the observation plane. These points are displayed using kernel density estimation (KDE), a nonparametric procedure that does not assume a prescribed functional form for the spatial distribution. In a KDE, each arrival point contributes a localized kernel to the estimated density, and the superposition of these contributions provides a smooth representation of the ray-arrival distribution. We use the \texttt{seaborn.kdeplot} routine with a Gaussian kernel. The resulting KDE plots avoid the explicit bin-size dependence of ordinary histograms and are interpreted here as smooth ray-density maps, not as wave-optical intensity distributions.

For clarity, the ray-tracing procedure can be summarized as a sequence of elementary operations:
\begin{itemize}
\item Step 1. Determine the intersection of the incident ray with the first surface.

\item Step 2. Compute the corresponding surface normal.

\item Step 3. Apply Snell's law to obtain the transmitted direction.

\item Step 4. Propagate the ray to the second surface.

\item Step 5. Repeat the refraction procedure at the second interface.

\item Step 6. Propagate the transmitted ray to the observation plane.
\end{itemize}
These steps define a compact and reproducible algorithm that follows directly from the equations above and can be implemented as a short computational project.


\section{Axicon-type geometries}
\label{sec:profiles}

The ray-tracing model of Sec.~\ref{sec:raymodel} can be applied to any pair of sufficiently smooth refracting surfaces. In this work we consider two families of axicon-type profiles: a conical profile and an exponential profile. The conical axicon provides the simplest geometry and gives a direct illustration of how axial symmetry produces ring-like ray distributions. The exponential-profile axicon introduces curved refracting surfaces and allows a richer dependence of the deflection angle on the impact parameter.

In both families, transverse scale parameters control the surface geometry along the $x$ and $y$ directions. Equal transverse scales on a given surface produce axial symmetry, whereas unequal scales introduce an elliptical deformation. By varying these parameters, one can therefore explore directly how symmetry breaking in the optical geometry transforms a ring-like ray distribution into multiple separated maxima, including the four-image patterns discussed below. This is the optical counterpart of the symmetry-breaking correspondence summarized in Table~\ref{tab:analogy}.


\subsection{Conical axicon}
\label{subsec:conical}

The first profile is the conical axicon, which is the standard axicon geometry in many optical applications. In the present model, the first surface is taken to be a plane parallel to the $X$$Y$ plane,
\begin{equation}
 z= f_1(x,y)=z_1,
 \label{eq:conical_f}
\end{equation}
where $z_1$ fixes the axial position of the planar entrance surface.
The second surface, on the other hand, is described by
\begin{equation}
 f_2(x,y) = z_2 \left[ 1 - \sqrt{\left(\frac{x}{\sigma_x}\right)^2 + \left(\frac{y}{\sigma_y}\right)^2 } \right] .
 \label{eq:conical_g}
\end{equation}
The parameter $z_2$ specifies the axial position of the conical apex, while transverse scales $\sigma_x$ and $\sigma_y$ determine the surface slopes, $z_2/\sigma_x$ and $z_2/\sigma_y$, along the $x$ and $y$ directions, respectively.

For $\sigma_x=\sigma_y=\sigma_2$, the surface describes a cone of revolution with circular symmetry, such that
\begin{equation}
 z= f_2(x,y) = z_2 \left( 1 - \frac{r}{\sigma_2}\right) ,
 \label{eq:conical_r}
\end{equation}
where $r = \sqrt{x^2 + y^2}$ denotes the radial (transverse) distance measured from the optical axis.
For $\sigma_x\neq\sigma_y$, the circular symmetry is replaced by an elliptical deformation of the cone (as seen from the front of the lens).
If the apex is laterally displaced from the optical axis to $(d_x,d_y)$, the centered coordinates $x$ and $y$ are replaced by $x-\delta_x$ and $y-\delta_y$, respectively:
\begin{equation}
 z=f_2(x,y) = z_2 \left[ 1 - \sqrt{\left(\frac{x-\delta_x}{\sigma_x}\right)^2 + \left(\frac{y-\delta_y}{\sigma_y}\right)^2 } \right] .
 \label{eq:conical_off}
\end{equation}
Thus, $\sigma_x=\sigma_y$ defines an axially symmetric conical axicon, whereas $\sigma_x\neq\sigma_y$ introduces an elliptical deformation of the exit surface.

This profile is useful because its effect on the rays is especially transparent. In the symmetric case, rays with the same radial distance from the optical axis experience equivalent deflections, and the arrival points on the observation plane form a ring-like pattern. In the asymmetric case, different transverse directions are refracted differently, which redistributes the ray density into separated ray-density maxima. Thus, even this simple profile illustrates how image morphology is controlled by the geometry of the refracting surface.

\begin{figure}[!t]
 \centering
 \includegraphics[width=\columnwidth]{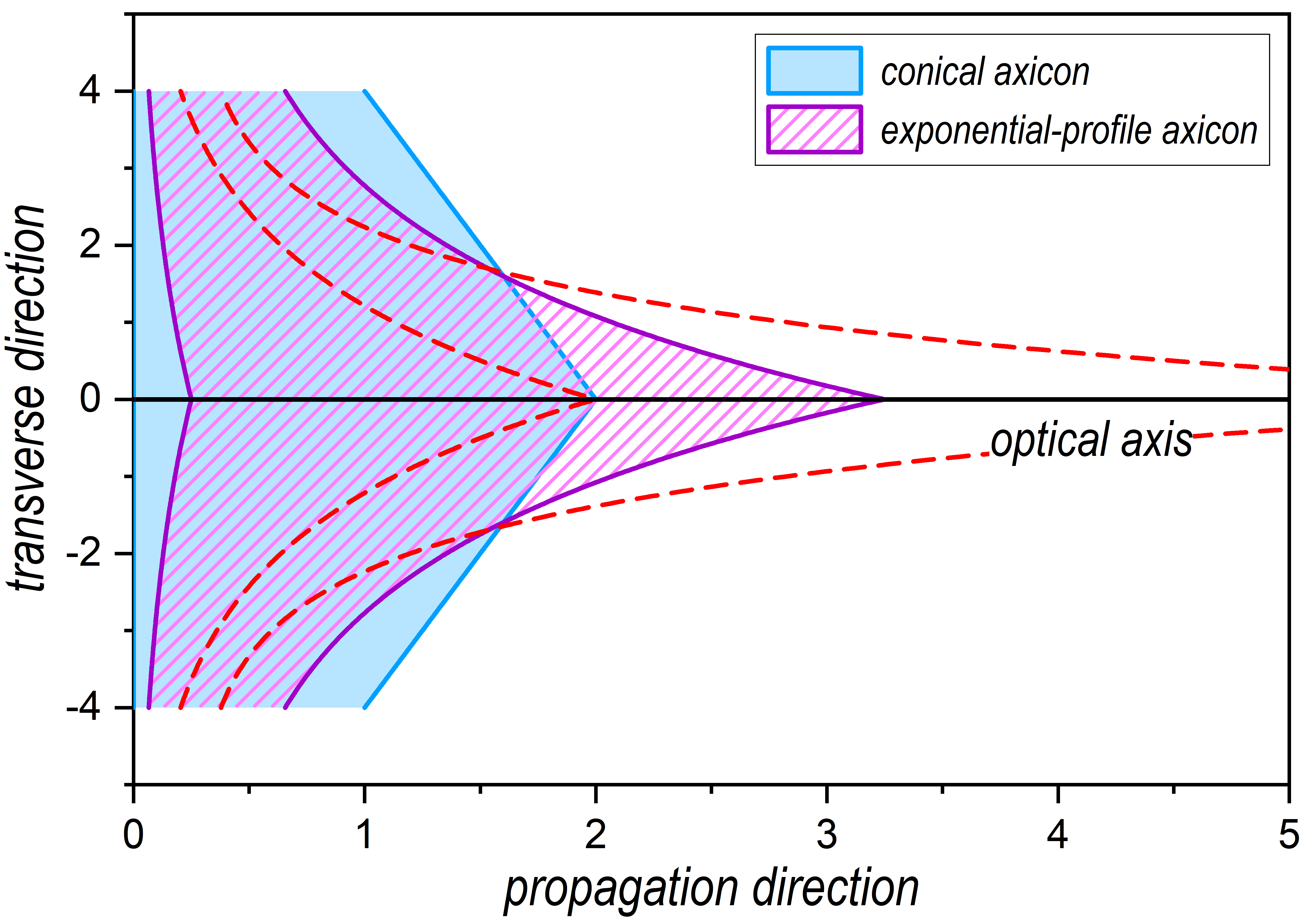}
 \caption{\label{fig2}
  Cross section of a conventional conical axicon (blue shaded area) and exponential-profile axicon (purple hatched area), both with radial symmetry around the optical axis.
  The parameter values for the conical axicon are: $z_1 = 0$, $z_2 = 2$, and $\sigma_2 = 8$.
  For the exponential-profile axicon, the set of parameter values are $f_1\!:\! \left\{ z_{\rm off,1} = 0, z_1 = 0.25, \sigma_1 = 3 \right\}$ and $f_2\!:\!\left\{ z_{\rm off,2} = 0.25, z_2 = 3, \sigma_2 = 2 \right\}$.
  The dashed red curves represent an exponential-profile axicon with a more pronounced apex, illustrating a more strongly curved optical profile.
  In this case, the parameters are  $f_1\!:\!\left\{ z_{\rm off,1} = 0, z_1 = 2, \sigma_1 = 1.75 \right\}$ and $f_2\!:\!\left\{ z_{\rm off,2} = 0.25, z_2 = 7, \sigma_2 = 1 \right\}$.}
\end{figure}

The blue shaded area in Fig.~\ref{fig2} shows a cross section of an axially symmetric conical axicon.
In the simulations below, this profile is used both to obtain ring-like patterns and to show how broken axial symmetry leads to the appearance of four-image configurations.


\subsection{Exponential-profile axicon}
\label{subsec:logarithmic}

The second family consists of axicon-type lenses with exponential radial profiles. In this case, both refracting surfaces are described by the functional form
\begin{equation}
 z=f_j(x,y)=z_{\mathrm{off},j}+z_j
 \exp\!\left[
 -\sqrt{
 \left(\frac{x}{\sigma_{x,j}}\right)^2+
 \left(\frac{y}{\sigma_{y,j}}\right)^2
 }
 \right],
 \label{eq:log_profile}
\end{equation}
where $j=1$ and $j=2$ label the front and rear refracting surfaces, respectively, each of which has its own set of parameters. The parameter $z_{\mathrm{off},j}$ specifies the asymptotic plane approached by the surface at large transverse distances, whereas $z_j$ gives the vertex height measured relative to that plane. Accordingly, the axial position of the vertex is
\begin{equation}
 f_j(0,0)=z_{\mathrm{off},j}+z_j.
 \label{eq:logarithmic_vertex}
\end{equation}
The transverse scales $\sigma_{x,j}$ and $\sigma_{y,j}$ control how rapidly the surface approaches its asymptotic plane along the $x$ and $y$ directions. Large values produce broader, more gently varying profiles, whereas small values produce narrower profiles with a more rapid variation near the optical axis. These transverse scales do not change the vertex height.

Figure~\ref{fig2} compares cross sections of the profiles introduced above. The blue shaded area represents the axially symmetric conical axicon discussed in Sec.~\ref{subsec:conical}. The purple hatched area represents an axially symmetric exponential-profile axicon with moderate transverse curvature, whereas the dashed red curves delimit a narrower exponential profile with a more pronounced apex region. The latter illustrates how decreasing the transverse scales $\sigma_{x,j}$ and $\sigma_{y,j}$ produces a more rapid variation of the surface near the optical axis. All three profiles are shown for representative optical parameters and are not fitted to an astrophysical mass distribution.

From a geometrical perspective, the exponential profile provides a variable-slope alternative to the conical axicon.
Whereas the conical surface has a constant slope away from the apex, the exponential profile has a local inclination that varies continuously with transverse distance from the optical axis. This difference ultimately produces the stronger dependence of the deflection angle on impact parameter observed for the exponential profile.

For each exponential-profile surface, axial symmetry is obtained when
\begin{equation}
 \sigma_{x,j}=\sigma_{y,j},
 \qquad j=1,2.
 \label{eq:logarithmic_axial_symmetry}
\end{equation}
The complete two-surface axicon is therefore axially symmetric provided that this condition holds independently at both interfaces; the transverse scales of the two surfaces need not be equal. If $\sigma_{x,j}\neq\sigma_{y,j}$ for either surface, that optical surface acquires an elliptical transverse profile. Allowing the two surfaces to have independent transverse scales provides additional control over the outgoing ray distribution and will be used below to obtain four separated ray-density maxima.

It is important to emphasize that the parameters introduced above are not fitted to any astrophysical lens. Their role is to control the optical geometry and to show, within a reproducible ray-tracing model, how changes in surface curvature and symmetry affect the resulting ray-density map. This is the sense in which the axicon functions as a geometrical analogue rather than as a physical model of a gravitational lens.


\section{Simulated lensing-like images}
\label{sec:results}

We now apply the ray-tracing procedure described in Sec.~\ref{sec:raymodel} to the axicon profiles introduced in Sec.~\ref{sec:profiles}.
The purpose of these simulations is not to reproduce a particular gravitational lens, but to illustrate how the geometry and symmetry of a simple optical element control the ray distribution on an observation plane.
In this regard, all spatial coordinates and geometrical parameters are expressed in the same arbitrary length unit.
The incident bundle consists of $N=40\,000$ parallel rays whose initial positions $(x_0,y_0)$ are sampled uniformly over a circular region of radius $R_0=5$ in the object plane at $z_0=0$.
Unless otherwise stated, the incident rays are all parallel to the optical axis ($\theta_0 = \varphi_0 = 0$).

At the observation plane $z=z_3$, the arrival points $(x_3,y_3)$ are displayed using the KDE procedure described in Sec.~\ref{sec:raymodel}. Brighter regions in the resulting maps indicate higher ray-arrival densities.

\subsection{Ring-like images from axially symmetric axicons}
\label{subsec:rings}

We first consider the axially symmetric conical confi\-gu\-ration defined in Sec.~\ref{subsec:conical}, with $\sigma_x = \sigma_y = \sigma_2$.
This symmetry is the optical analogue of the axial symmetry required, in an ideal gravitational-lensing configuration, for the formation of an Einstein ring. In the present model, the axial symmetry of the refracting surfaces ensures that rays launched at the same radial distance from the optical axis are deflected in equivalent ways. As a result, the ray density on the observation plane is concentrated around a ring-like structure.

\begin{figure}[!t]
 \centering
 \includegraphics[width=\columnwidth]{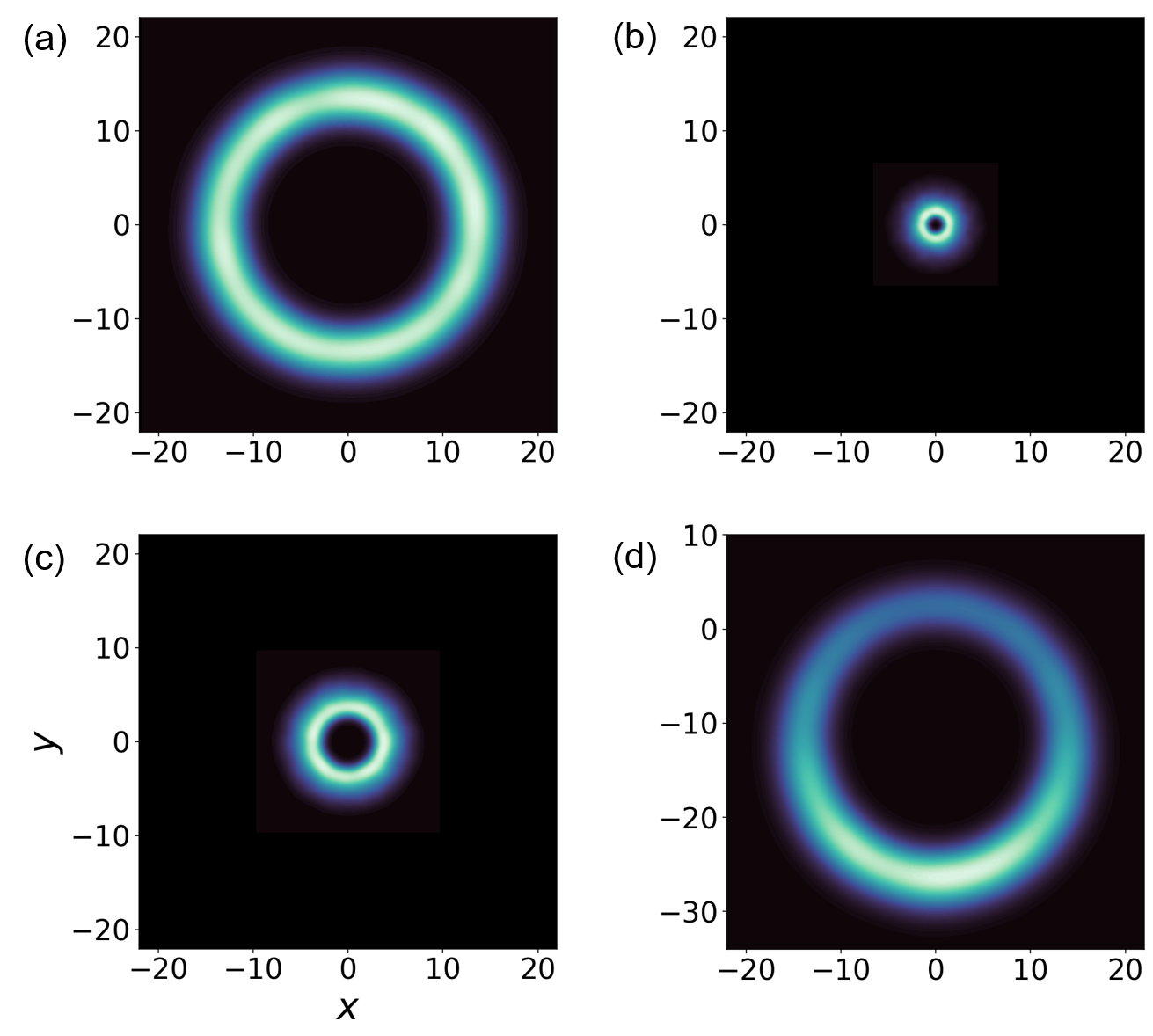}
 \caption{\label{fig3}
  Ray-density maps obtained with a conical axicon with refractive index $n=1.5$, surrounded by a medium with unit refractive index, and with parameters: $z_1 = 10$, $z_2=25$, and $\sigma_2=60$.
  The object is represented by $N=40,\,000$ rays with initial points randomly uniformly distributed over a circular region of radius $R_0=5$ in the object plane, and all launched parallel to the optical axis ($\theta_0 = \varphi_0 = 0$).
  (a) Axially symmetric ring-like structure observed on a plane at $z_3=100$. 
  (b) Axially symmetric ring-like structure observed on a plane at $z_3=50$. 
  (c) Axially symmetric ring-like structure observed on a plane at $z_3=100$ for $\sigma_2=120$, illustrating how surface geometry controls the ring radius.
  (d) Broken axial symmetry by launching the rays with an initial tilting ($\theta_0=2\pi/45, \varphi_0=3\pi/2$), which gives rise on the observation plane to an arc-type structure.}
\end{figure}

Figure~\ref{fig3} shows representative results for a conical axicon with refractive index $n=1.5$, surrounded by a medium with unit refractive index, and determined by the parameters: $z_1 = 10$, $z_2=25$, and $\sigma_2=60$.
In Fig.~\ref{fig3}(a), we find that the image observed on a plane with $z_3=100$ has a ring-like ray-density structure with axial symmetry.
The geometrical ray mapping predicts an outer radius $R_{\rm out}\simeq16.7845$, associated with rays launched arbitrarily close to the optical axis, and an inner radius $R_{\rm in}\simeq12.2507$, associated with rays launched from the boundary of the incident circular bundle.
The image points are thus inverted with respect to their positions at the object plane.

For the situation represented in Fig.~\ref{fig3}(a), we find that the radial distance between the inner and outer boundaries of the ring, $\Delta R \approx 4.5338$, is relatively small compared to the inner diameter, $2R_{\rm in} \approx 24.50$.
If the observation plane is moved closer to the axicon, for example, to $z_3=50$, we obtain a much smaller annular distribution with a reduced central opening, as seen in Fig.~\ref{fig3}(b), although the distance between the outer and inner radii still remains the same.
Although the radial width remains unchanged, the inner and outer radii decrease to $R_{\rm in}\simeq1.061$ and $R_{\rm out}\simeq5.595$. Consequently, the distribution appears as a broad annulus with a small central opening rather than as a sharply defined thin ring.
An analogous situation takes place if the value of $\sigma_2$ is changed instead.
In Fig.~\ref{fig3}(c), $z_3=100$ but $\sigma_2=120$, and we have $R_{\rm out} \approx 7.9436$ and $R_{\rm in} \approx 3.0539$, which render $\Delta R \approx 4.8897$.
Increasing $\sigma_2$ reduces the surface slope $z_2/\sigma_2$ and therefore decreases the angular deflection of the transmitted rays.

\begin{figure}[!t]
 \centering
 \includegraphics[width=.8\columnwidth]{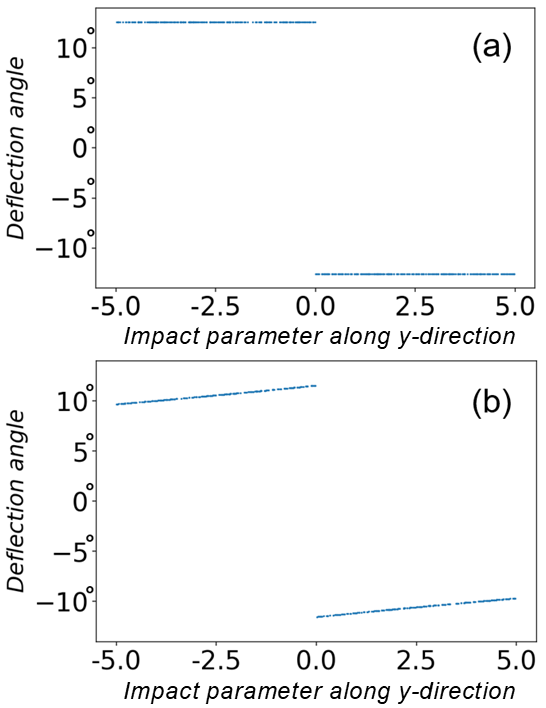}
 \caption{\label{fig4}
  Final deflection angle, defined as the angle between $\hat{\bf k}_0$ and $\hat{\bf k}_2$, as a function of the impact parameter along the $y$-direction for: (a) a conical axicon and (b) an exponential-profile axicon.
  In both cases, the axicon has refractive index $n=1.5$, has revolution symmetry around the optical axis, and is surrounded by a medium with unit refractive index.
  Parameters used with the conical axicon: $z_1 = 10$, $z_2=25$, $\sigma_2=60$, $z_3 = 100$.
  Parameters used with the exponential-profile axicon: $z_{\rm off,1} = z_{\rm off,2} = 0$, $z_1=30$, $z_2=40$, $\sigma_1=40$, and $\sigma_2=100/3$.
  The impact parameter is here defined as the $y$-component of $N=40,\,000$ initial points randomly uniformly distributed over the circular region of radius $R_0=5$ in the object plane, all launched parallel to the optical axis ($\theta_0 = \varphi_0 = 0$).
  Note that, while the conical profile gives constant deflection away from the apex, the exponential profile produces a deflection that varies continuously with the initial position (decreasing in modulus towards higher values of the radial position).}
\end{figure}

An additional feature of the mapping is that the image is effectively inverted: rays originating from the outer regions of the object contribute to the inner part of the ring, while rays originating closer to the center of the object contribute to the outer part. This behavior follows directly from the relation between impact parameter and angular deflection, as it can be seen in Fig.~\ref{fig4}(a), where the final deflection angle is plotted against the impact parameter along the $y$-direction (here defined as the $y$-component of the initial sampling points randomly uniformly distributed over the circular region of radius $R_0=5$ in the object plane).
The exponential-profile axicons introduced in Sec.~\ref{subsec:logarithmic} provide a useful contrast.
In this case, both refracting surfaces may be curved, and the local slope varies continuously with distance from the optical axis. Consequently, the final deflection angle is no longer approximately constant as a function of impact parameter, as shown in Fig.~\ref{fig4}(b). Compared with the conical case, the exponential-profile geometry produces a richer redistribution of rays and a stronger sensitivity to the local curvature of the refracting surfaces.
Indeed, in some configurations, the curvature is sufficiently strong that total internal reflection occurs for part of the incident ray bundle, generating gaps in the transmitted distribution.


\subsection{Partial arcs from misaligned incidence}
\label{subsec:arcs}

In gravitational lensing, a complete Einstein ring requires a high degree of alignment between the source, the lens, and the observer.
If this alignment is broken, the observed morphology may consist of incomplete ring segments or arcs.
This symmetry breaking is represented in the optical model by tilting the incident ray bundle.
The tilt is introduced through the polar and azimuthal angles $\theta$ and $\varphi$ in Eq.~\eqref{eq:k0}.
The conical example in Fig.~\ref{fig3}(d) shows that a nonzero incidence angle breaks the circular symmetry of the arrival distribution and produces an arc-like pattern.

A simple conical axicon thus shows how the arc-like pattern is controlled mainly by the global deflection angle and the observation distance. More generally, curved surfaces can introduce additional variations in the local ray density. This distinction provides a useful way of separating symmetry breaking due to incidence direction from redistribution due to surface curvature.


\subsection{Four-image configurations from broken axial symmetry}
\label{subsec:cross}

We next consider the asymmetric configurations introduced in Secs.~\ref{subsec:conical} and \ref{subsec:logarithmic}, in which the axial symmetry of one or both refracting surfaces is broken by choosing unequal transverse scales.
This deformation is the optical analogue of replacing an axially symmetric gravitational lens by an elongated or otherwise non-axisymmetric mass distribution. In gravitational lensing, such a configuration can produce multiple images of the same source, including the well-known Einstein-cross morphology. In the present optical analogy, the corresponding effect is the appearance of several separated maxima in the transmitted ray-density map.

\begin{figure}[!t]
 \centering
 \includegraphics[width=\columnwidth]{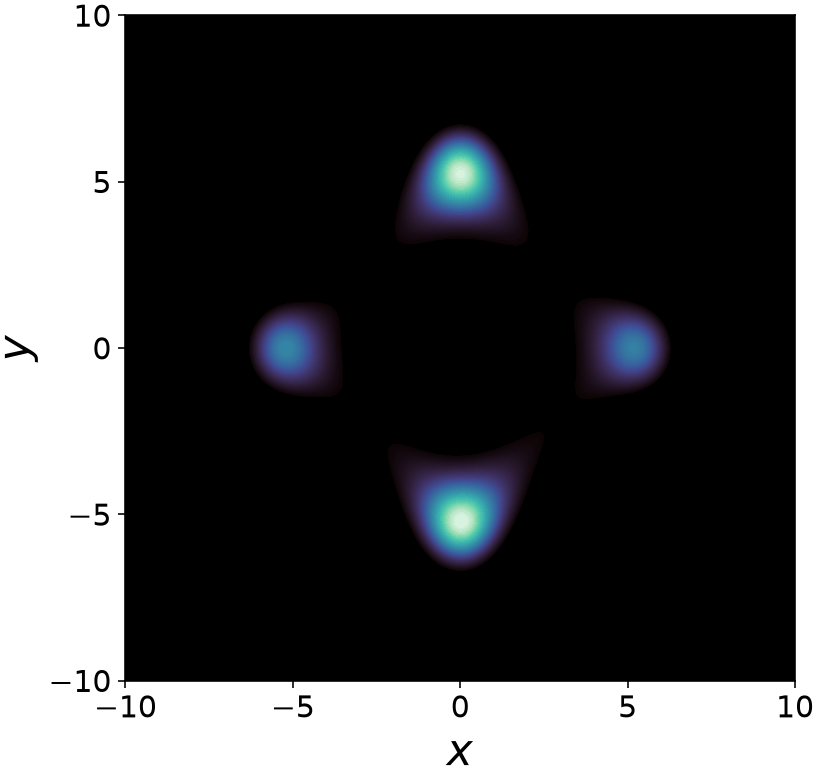}
 \caption{\label{fig5}
  Ray-density map obtained with an exponential-profile axicon with refractive index $n=1.4$, surrounded by a medium with unit refractive index, and with parameters: $z_{\rm off,1} = z_{\rm off,2} = 0$, $z_1=16$, $z_2=21.5$, $\sigma_{x,1}=9.88$, $\sigma_{y,1}=11.8$, and $\sigma_{x,2}=\sigma_{y,2}=\sigma_2=12.95$.
  The object is represented by $N=40,\,000$ rays with initial points randomly uniformly distributed over a circular region of radius $R_0=0.8$ in the object plane, and all launched parallel to the optical axis ($\theta_0 = \varphi_0 = 0$).
  The object plane is at $z_0=0$, and the observation plane is at $z_3=150$.
The unequal transverse scales of the front surface break the axial symmetry, while the independent curvatures of the two interfaces provide additional control over the outgoing ray distribution.
Together, these geometrical features redistribute the rays into four separated ray-density maxima, producing a pattern reminiscent of an Einstein-cross morphology.}
\end{figure}

Asymmetric conical axicons can already produce separated ray-density maxima, but the resulting pattern is rather sensitive to the observation distance and to the transverse scales $\sigma_x$ and $\sigma_y$.
The exponential profile offers additional freedom because the entrance and exit surfaces can have different curvatures and transverse asymmetries. Figure~\ref{fig5} shows a representative configuration that produces four well-separated maxima. This example is especially useful for illustrating the role of surface geometry: the entrance and exit surfaces do not contribute in the same way, and small changes in curvature or asymmetry can substantially modify the final ray-density pattern.

The four maxima in Fig.~\ref{fig5} are the result of a controlled redistribution of rays by an asymmetric optical geometry. Their value lies in showing, within a simple ray-tracing model, how breaking axial symmetry can transform a ring-like distribution into a set of separated ray-density maxima. This is the qualitative feature that motivates the comparison with an Einstein-cross morphology.

\section{Pedagogical use and limitations of the analogy}
\label{sec:pedagogy}

The model presented in the previous sections is intended primarily as a pedagogical and computational analogy. Its main value is that it allows students to explore, within the familiar framework of geometrical optics, how ray deflection and image morphology depend on geometry and symmetry. The ingredients of the model are elementary: parametric ray equations, surface normals, Snell's law, and propagation to an observation plane. Nevertheless, their combination produces nontrivial ray-density patterns that resemble some of the most recognizable morphologies of strong gravitational lensing.

A first possible use of the model is in an advanced undergraduate course on geometrical optics. In this context, the axicon provides an example that goes beyond the standard thin-lens approximation. Students can compare ordinary focusing by spherical lenses with the extended focal region of an axicon, and then examine how this different focusing behavior leads to ring-like ray distributions. The conical axicon is especially useful for this purpose because its constant slope leads to a simple relation between geometry and deflection. The exponential-profile axicon provides a natural extension in which the local surface slope varies with the impact parameter, giving rise to a more structured redistribution of rays.

A second use is in courses or projects involving computational physics.
Since the algorithm follows directly from Eqs.~\eqref{eq:r0}--\eqref{eq:Po}, it can be implemented as a short computational project.
For each ray, one finds the intersections with the two refracting surfaces, computes the corresponding normals, applies Snell's law, and propagates the transmitted ray to an observation plane. Students can then investigate how the final ray-density map changes when one varies the refractive-index contrast, the observation distance, the curvature of the surfaces, or the transverse symmetry parameters. In this way, the model provides a concrete setting for connecting mathematical equations with numerical visualization.

A third use is as an introduction to the qualitative language of gravitational lensing. The model can be used to discuss why axial symmetry favors ring-like images, why misalignment breaks a ring into arcs, and why broken axial symmetry can lead to several separated ray-density maxima. These ideas are central to the interpretation of strong-lensing images, but here they can be approached without first deriving the full gravitational lens equation. The analogy therefore provides a useful intermediate step between elementary geometrical optics and more advanced treatments of gravitational lensing.

The limitations of the analogy should be stated explicitly. The axicon does not reproduce a space-time metric, and the rays inside the optical element are not null geodesics of a gravitational field. The parameters of the axicon are not related to a lensing mass distribution, and no attempt is made to reproduce realistic angular scales, magnifications, time delays, flux ratios, or caustic structures of a particular gravitational-lensing system. The output images should therefore be interpreted only as qualitative ray-density patterns. Their similarity to Einstein rings, arcs, and Einstein-cross-like configurations is morphological and pedagogical, not dynamical or astrophysical.

This limitation is also what distinguishes the present construction from approaches in which a gravitational field is represented through an effective, spatially varying refractive index of the vacuum. Here the refractive index is constant within each region, and the redirection of rays is produced only by the shape of the interfaces. Thus, the curved optical surfaces are not a physical representation of curved space-time, but a geometrical device for redistributing rays in a controlled and reproducible way.

Another limitation concerns the use of ray density as a proxy for image intensity. The simulations presented here are based on geometrical optics and do not include diffraction, interference, polarization, finite aperture effects, or detector response. In regions where rays cross or become highly concentrated, a wave-optical treatment would be required to describe the actual optical field. For the purposes of the present analogy, however, the ray-density maps are sufficient because the goal is to visualize the geometrical origin of the image morphologies rather than to predict physical intensities.

Making these limitations explicit is part of the pedagogical value of the model.
A clear distinction between what is included and what is omitted helps students distinguish between a qualitative or computational analogy and a physical theory.
It also provides an opportunity to discuss the role of idealization in physics: a simplified system may fail to reproduce the full dynamics of a phenomenon while still capturing selected structural features that are useful for understanding.


\section{Conclusions}
\label{sec:conclusions}

We have presented a geometrical-optics analogy for gravitational lensing based on ray tracing through axicon-type lenses. The central idea is to use optical elements with constant refractive index and to let the geometry of their refracting surfaces determine the redistribution of rays. This approach is complementary to analogies based on graded refractive indices: instead of encoding the effect in a spatially varying optical medium, the present model emphasizes the role of surface geometry and symmetry.

The ray-tracing formulation was developed for a general two-surface axicon and then applied to conical and exponential profiles. For the conical profile, the constant surface slope leads to a simple and transparent deflection pattern. For the exponential profile, the varying surface inclination introduces a dependence of the deflection on the impact parameter, producing richer ray-density distributions. Both profiles can be treated within the same computational framework using only elementary tools from geometrical optics.

The simulations illustrate three qualitative correspondences with gravitational-lensing phenomenology. Axially symmetric axicons generate ring-like ray-density maps, providing an optical analogy of an Einstein ring. Tilting the incident ray bundle breaks the circular symmetry of the image and produces partial arcs, mimicking the effect of source-lens-observer misalignment. Finally, breaking the axial symmetry of the axicon by using unequal transverse scale parameters can produce several separated ray-density maxima, reminiscent of an Einstein-cross morphology.

Rather than introducing a new optical simulator of gravitational lensing, the present work provides a compact and reproducible framework for studying how surface geometry and symmetry redistribute light rays into ring-like, arc-like, and multiple-maximum patterns. Its explicit assumptions and elementary geometrical-optics ingredients make it suitable for advanced undergraduate teaching and computational physics projects. More broadly, the construction illustrates how a carefully delimited analogy can isolate a structural feature of a complex phenomenon and make it quantitatively accessible without being mistaken for a physical equivalence.


\section*{Data Availability Statement}

The data that support the findings of this study are available from the corresponding author upon reasonable request. The geometrical and optical parameters used in the reported simulations are provided in the article.



%

\end{document}